\documentclass[twocolumn,showpacs,preprintnumbers,amsmath,amssymb,nofootinbib
,superscriptaddress]{revtex4}
\usepackage{hyperref}
\usepackage{graphicx}
\usepackage{color}
\usepackage{xcolor}

\usepackage{diagbox} 
  
\def\beq{\begin{equation}}  
\def\eeq{\end{equation}}  
\def\bea{\begin{eqnarray}}  
\def\eea{\end{eqnarray}}  
\def\half{\frac{1}{2}}  
  
\def\bq{\begin{quote}}  
\def\eq{\end{quote}}

\def\GeV{\,{\rm GeV}}       
\def\eV {\,{\rm  eV}}

\def\half{\frac{1}{2}}       

\relax

\begin{document}
\preprint{FERMILAB-PUB-19-457-T}
\preprint{IFIC/19-37}

{
\title{ Sterile Neutrinos, Black Hole Vacuum \\ and Holographic Principle  }
\author{Gabriela Barenboim}
\affiliation{Departament de F\'{\i}sica Te\`orica and IFIC, \\
Universitat de 
Val\`encia-CSIC, E-46100, Burjassot, Spain.}
\author{Christopher T. Hill}
\affiliation{Theoretical Physics Department, \\
Fermi National Accelerator Laboratory, \\
P. O. Box 500, Batavia, IL 60510, USA}

\begin{abstract}
We construct an effective field theory (EFT) model that describes matter field interactions
with Schwarzschild mini-black-holes (SBH's), treated as a scalar field, $B_0(x)$.
Fermion interactions with SBH's require
a complex spurion field, $\theta_{ij}$, which we interpret as the EFT description
of ``holographic information,''  which
is correlated with the SBH as a composite system.
We consider Hawking's virtual black hole vacuum (VBH) as a Higgs phase, $\langle B_0 \rangle =V$.  
Integrating sterile neutrino loops, the information field $\theta_{ij}$ is promoted to a dynamical field, 
necessarily developing a tachyonic instability and acquiring a VEV of order the Planck scale.
For $N$ sterile neutrinos this  breaks the vacuum to  $SU(N)\times U(1)/SO(N)$ 
with $N$ degenerate Majorana masses, 
and $\frac{1}{2}N(N+1)$ Nambu-Goldstone neutrino-Majorons.
The model suggests
many scalars fields, corresponding to all fermion bilinears, may exist bound nonperturbatively
by gravity.
\end{abstract}

\maketitle
 
\date{\today}


\section{Introduction}

In the present paper we will discuss the issue of the ``black hole information paradox'' 
in the context of an effective quantum field theory (EFT). 
Classical EFT is a powerful
tool for summarizing low energy processes where the detailed short-distance behavior of a
system is integrated out. Examples of its application to black holes can be found in  \cite{Wong:2019yoc} and references therein, where effective field theories are world-line actions and carry local operators that represent known cases of the breakdown of classical no hair theorems. We also implicitly rely on the intuition of Dvali and Gomez in their picture
of threshold quantum black holes \cite{DG}.

We assume there is a quantum limit for black holes, near the threshold of
production or collisions in scattering processes, or in coherent processes such as Bose-Einstein condensation.  These processes can exist in field theoretic descriptions even through
the underlying objects are complicated many body states, {\em e.g.,} Rubidium atoms with $Z=37$
can be described as pointlike objects, yet form a quantum condensate which can be
described as the vacuum value of a field. Hadronic boundstate production can be
described by pointlike interactions despite the short distance complexity of a hadron.

Presently we will decribe
a black hole by a field, $B(x)$, which can create or destroy a black hole of mass $M$
at the event $x$ and satisfies a free field equation of motion.  We ignore the
warped external geometry at large distances compared to the black hole Schwarzschild radius of $B(x)$ (following \cite{Wong:2019yoc}). 
Fundamental problems must then be faced in constructing the interactions of
$B(x)$ with matter.  

We consider the simplest case of a pair of scalars
which possess a global $SO(2)=U(1)$ symmetry.  How do we write down a
pointlike interaction of a pair of scalars $\phi_i$ and $\phi_j$ with a 
real Schwarzschild black hole $B(x)$? 
By ``no-hair'' reasoning any
interaction, such as  $\phi_1+\phi_1\rightarrow B$ must have the same physical
rate, or same quantum amplitude, as any other interaction
such as $\phi_1+\phi_2\rightarrow B$.  A naive ``no-hair''
theorem would say these must be indistinguishable processes.

However, it is not possible to
mathematically write down interactions that would have this universality.
Once we commit to a certain amplitude for $\phi_i+\phi_j\rightarrow B$
then we will generally not have an equivalent amplitude for some other process
$\phi'_i+\phi'_j\rightarrow B$ where $(\phi'_i,\phi'_j)$  is a general linear
superposition of the $(\phi_i,\phi_j)$.
This is only possible in special cases, such as if we maintain 
the symmetry $U(1)$ and restrict ourselves
to $U(1)$ transformations of the fields, 
in which case the process $\phi_1+\phi_1\rightarrow B$ is
forbidden.

Stemming from this it is conventional wisdom that the global
symmetry is broken by quantum gravity. Then different configurations
of the two scalars may have different amplitudes to interact with $B(x)$.
However, given that both
processes 
$\phi_1+\phi_2\rightarrow B$ and $\phi_1+\phi_1\rightarrow B$ must exist,
since gravity is ``flavor blind,''
what rule dictates their amplitudes?  The breaking of  a symmetry
is actually expressed by the form of the interaction.  Breaking a symmetry
actually {\em requires that we supply more information, not less.} 

One might argue that the quantum effective field theory does not exist,
and the S-matrix does not exist, at Planckian high energies. 
However, given a mini black hole, we can consider an idealized low energy experiment
where the resolvable distance scales are large. 
We make an arbitrary initial combination of the two $\phi_i$'s and
allow them to collide with the mini Schwarzschild black hole, $B$, resulting in a final
state black hole $B'$. We take pains to observe the exclusive process
where no other particles appear in the initial or final states.
Our question is, what is the effective
interaction $g\theta^{ij}\phi_i\phi_j B B'$?   What then dictates
the couplings $\theta_{ij}$?  Theoretically or experimentally 
determining $\theta_{ij}$
may be problematic, but we cannot deny its existence.

 Quantum mechanics 
conflicts with no-hair theorems. If we believe in  Dirac's
formulation of the Hilbert space of superpositionable states,
an S-matrix and Wilsonian effective field theory (which we do) then
we must
allow for the possibility that $B(x)$ carries information. 
To us the resolution is that the information contained in the $\theta^{ij}\phi_i\phi_j$
initial state is transfered to the $B'$ state, and resides on it's horizon.
Essentially the black hole $B'$ has acquired ``information hair'' and is
now defined by a field $B'=\theta^{ij} B(x)$ . This acquired information supplements
any other information that was already present on  $B$, though not probed
in the experiment. 

In Section II we will reiterate these issues in more detail
for a pair of scalar fields $\phi_i$, $i=(1,2)$.  
 
To connect to real-world fields and make these issues more concrete
we  consider, in addition to scalars, sterile neutrinos, possessing
an $SU(N)\times U(1)$ global symmetry, and consider how they interact
with a Schwarzschild mini-black hole (SBH) of mass $M_{Planck}$. 
In an effective field theory, 
we describe the SBH by a real quantum field, $B_0(x)$,  
whose excitations are minimal mass, 
tiny black holes.  We can follow the intuition of
\cite{DG}, who view the spectroscopy
of mini-blackholes as a tower of states labelled by a quantum number
$N$ where $N$ is effectively the number
of self gravitating gravitons in the core
of the BH.

We know that an $s-$wave pair of sterile neutrinos can
fall into the SBH, or scatter close to it.
We can write couplings of the neutrinos to $B_0$ that preserves or violates the
global symmetry. This then raises the question of how the 
neutrino number current is carried (or destroyed) by the black hole?

A conserved global-charge current cannot be carried by a real scalar field alone. 
To engineer
a coupling of a Weyl neutrino pair, $\sim \epsilon^{\alpha\beta} \nu_\alpha^i(x)\nu_\beta^j(x)$, to $B_0(x)$ 
we need a complex ``information'' that is either
intrinsic to the black hole or is a field 
that can be attached to the SBH.  We will designate this as $\theta_{ij}(x)$.  
Physically, the neutrino pair falls into the SBH, but never appears to cross the horizon. 
However, at some point we  can no longer distinguish the neutrino pair and black hole from
a pure black hole,
though there may be some information record 
in the EFT black hole that is  absorbing them.  

In the effective field theory
we describe the information by a ``spurion,''
a complex field that transforms under $SU(N)$ identically to the neutrino pair
and allows us to tie all indices together in the interaction vertex.
This is co-localized 
with the SBH and the neutrinos at the interaction vertex. The spurion represents
information on the horizon of the SBH.   The EFT is describing a mini-black hole
and the region immediately surrounding it, external to the horizon, which includes anything orbiting or in the
process of infall as seen by the Schwarzschild observers
and this includes $\theta_{ij}(x)$.

We distinguish three possible logical cases for a dynamical $\theta_{ij}$. 
(I) Information is explicitly lost;  (II) Information is conserved but does not propagate;  
(III) Information is conserved
and is carried by the black hole.  

Case (I) evidently implies an explicit, arbitrary fixed value of the spurion permeating all of space
with a fixed orientation in the group space, hence it breaks $SU(N)\times U(1)$ transformation, 
and the associated Noether current is violated.  It seems that this case makes no sense
fundamentally, since there
is no procedure to specify $\theta_{ij}(x)$, even as a random variable,
however it may arise spontaneoeuly through the formation of
 a condensate of underlying black holes.
Note that just naively summing indices,
$\Sigma_{ij} \epsilon^{\alpha\beta} \nu_\alpha^i\nu_\beta^j B_0(x)$, implies a particular basis choice and therefore
a particular choice of $\theta_{ij}(x)\sim \Sigma_{nm} \delta_{in}\delta_{jm}$, a matrix with all  elements $=1$.  This is logically equivalent to case I and seems unreasonable
unless it arises by spontaneous symmetry breaking.

Case (II) implies that we introduce a random field variable  $\theta_{ij}(x)$, that has no correlation with
the black hole itself and, minimally, has no derivatives.  This implies that there is no current associated with
 $\theta_{ij}(x)$, and information is not carried by the SBH, yet the neutrino global current is conserved.
We can treat $\theta_{ij}(x)$ as a random field, and average over $\theta_{ij}(x)$
configurations through which the black hole and the neutrinos
propagate. This is somewhat akin to a ``spin-glass'' in condensed matter physics \cite{Edwards}. It simply
 ends up promoting   $\theta_{ij}(x)$ to being a propagating dynamical field. 

Case (III) implies that the information is attached to the effective SBH and is
dynamically transported by it. This requires a``conjoined kinetic term'' where the  $\theta_{ij}(x)$
moves with $B_0(x)$ as a composite field, $\sim \theta_{ij}(x)B_0(x)$. Here $\theta_{ij}(x)$ 
is positionally correlated with $B_0(x)$.  Case (III) is the most
sensible to us, and we interpret it tentatively as an {\em effective field theory
description} of the holographic principle \cite{holographic}.  
Once we engineer the interaction described in Case (III) above, we find that there
are potential dynamical consequences. 

As an application, we consider the effect of this on the neutrino vacuum.
The vacuum of the underlying $B_0(x)$ field  may be nontrivial.
Indeed, we know of three classical and important (effective) scalar
fields in nature, the Higgs field of the standard model (SM), the $\sigma$-meson of QCD
chiral dynamics, and the Ginzburg-Landau field of a superconductor (an effective 
description of a Cooper pair).
In each of these examples the vacuum is a ``Higgs phase.'' In our present scenario we
consider the possibility that $B_0(x)\rightarrow V +{B}$ where $V$ is a nontrivial
vacuum expectation value (VEV). 

A black hole Higgs phase will, in our model, have implications for the dynamical
behavior of information encoded in $\theta_{ij}$: owing to the effects
of neutrino loops external to the horizon (in the EFT)
 {\em global information becomes a propagating massless
field in a Higgs phase of $B_0(x)$}.   Moreover, this back reaction of the neutrino fields
induces an instability, causing a tachyonic potential
for $\theta_{ij}(x)$ to develop, and in turn,  $\theta_{ij}(x)$ acquires a VEV. 
This happens in both cases (II) and (III) (though there are slight differences) and
we obtain an effective potential that leads to a VEV
for $\theta_{ij}$, and sterile  neutrinos  develop Planck scale masses.
The $SU(N)\times U(1) $ global symmetry is broken to $SO(N)$.

\section{Quantum Inconsistency with the ``No-Hair'' Theorem}

Consider the ``no-hair'' theorem of classical black holes. 
This allows for a black-hole, $B$, to have gauge charges, or
quantum numbers. For example, black holes with the quantum numbers
 of the standard model Higgs boson could in principle be produced 
in collisions, such as a pair of electrons $e_L + \bar{e}_R\rightarrow B$.
However we could equally well consider a collison
$c_L + \bar{t}_R\rightarrow B$ of a 
lefthanded-charm quark with an anti-righthanded-top quark. 
These initial states have the same gauge charges, but involve different
combinations of flavors of initial state fermions.
Gravity, being flavor blind, supposedly
cannot make a flavored black hole.
Hence, $B$ must be  universally
coupled to any and all fermion bilinears.

However, a universal coupling to any and all fermion bilinears cannot
exist in quantum mechanics.
In the above example if we allow the latter process 
then there will be some mixed state collision,
such as a Cabibbo rotated charm-strange combination, $(c_L\cos\theta + s_L\sin\theta)+ \bar{t}_R\rightarrow B$,
{\em where we did not Cabibbo rotate the top quark}.
This process will have a $\theta$-dependent rate
and thus does not respect the flavor-blind universality
of gravity. 

To simplify, 
consider the production of a black hole in a collision of
two scalar particles
$\phi_i$ , with two global flavors, $i=1,2$.  
We want to describe the process
$\phi_i + \phi_j \rightarrow B$ in an effective field theory.
We could 
describe this process directly by an $S$-matrix element,   $\langle ij|B\rangle $ where
$B$ can be a Schwarzschild black hole
if the collision of the scalars is $s$-wave.   The existence
of the $S$-matrix implicitly assumes
that $|B\rangle $ is a quantum state.

Since gravity is flavor blind, a no-hair theorem
would evidently imply that the $S$-matrix element,    $\langle ij|B\rangle =S $,
must be a constant independent of the choice of normalized in-states for the $\phi_i$ particles,
for arbitrary $i$ and $j$. (A weaker statement might
be that the probability is the same for each such initial state).
In a quantum field theory description the S-matrix must exist
and we can introduce a field
for the black hole which we designate as $B(x)$,
and we also have the two fields $\phi_i(x)$.
Our equivalent field theory problem is then,
how do we write the local interaction vertex for 
the process $\langle ij|B\rangle $?

Let us introduce a complex field $\Phi = (\phi_1+i\phi_2)/\sqrt{2}$
Consider the  interaction:    
\beq
 g|\Phi|^2B = \half g \Sigma_i \phi^2_i  B 
\eeq
This would forbid   $\phi_1 + \phi_2  \rightarrow B$
 {\em i.e.,}, $i\neq j$. 
It equivalently forces the $S$-matrix
to be diagonal in flavor, $\langle ij|B\rangle =\delta_{ij}S $
hence $ \langle 1,2|B\rangle =0$.  Therefore a $U(1)= 
SO(2)$ singlet cannot decribe the process since we certainly expect that $\phi_1+\phi_2$ can
collide to make a Schwarzschild black hole and should have the same 
S-matrix.  

However, consider the interaction vertex,     
\beq
 ig (\Phi^2-\Phi^{*2}) B 
 =-2g \phi_1 \phi_2 B
\eeq
This breaks the $U(1)$ symmetry and 
indeed now describes the process
$ \langle 1,2|B\rangle =2g$.
However, suppose we consider the collision $ \langle 1',2'|B\rangle =2g$
where:
\bea
|1'\rangle  &=&  \cos(\theta) |1\rangle + \sin(\theta) |2\rangle
\nonumber \\
|2'\rangle  &=&  -\sin(\theta) |1\rangle + \cos(\theta) |2\rangle
\eea
These are simple superimposed states and can readily be produced in a laboratory,
such as with neutral K-mesons, or oscillating neutrinos.  However,
we now see that the amplitude is theta dependent:
\bea
\langle 1',2'|B\rangle = 2g \cos(2 \theta)
\eea
Hence we cannot write a local field description of  $\phi_i + \phi_j  \rightarrow B$
that yields an $S$-matrix, $\langle ij|B\rangle =S $  that is constant and independent of 
$i$ and $j$.  The flavor blindness or
``no-hair'' theorem, in this sense, is incompatible with quantum theory.

The only sensible resolution  is that the black hole does have
memory of the initial state, with global quantum numbers. The
black hole field is then  $\sim B^{ij}$ which
is a BH that has quantum flavor information, hence "hair".  
The vertex takes the form
\bea
g\Sigma_{ij}  \phi_i \phi_j B^{ij}
\eea
$B^{ij}$ remembers  the initial states that made it and the resulting
interaction is $SO(2)$ invariant.  Equivalenly, the black hole
must be a complex field in the interaction $\Phi^2 B + hc$ and 
the phase transformation $\Phi\rightarrow e^{i\theta}\Phi$ together
with $B\rightarrow e^{-2i\theta}B$ must be a symmetry.
This implies the BH carries a global $U(1)$ charge.

By insisting on the classical
no-hair theorem we would conclude that global charges cannot exist.
However, it has been recently understood that classically time dependent hair exists.
For example, reference \cite{Wong:2019yoc}
nicely treats and reviews the situation.
A BH immersed in a time dependent scalar field will absorb and
reemit this field producing a time dependent, coulombic ``halo.'' 
If, for example, the time dependent field is a Nambu-Goldstone boson
(eg, axion) then the associated charge is given by the chiral current $f \partial_t\phi$
so indeed, the BH has acquired a global chiral charge density.

Moreover, by claiming that the $U(1)$ symmetry is 
broken by gravity, by itself, in no way resolves this issue. Even if
global symmetries do not exist at short distances, but quantum
mechanics applies at all distances, then we face these issues. The Dirac superposition
of Hilbert space states is the underling conflict here. In this regard, we should
mention that
`t Hooft has reformulated
quantum mechanics with an underlying ``digital,'' cellular automata structure  \cite{Hooft:2014kka} 
where the Dirac superposition of states is only effective, but
not fundamental. Presumably Dirac superposition arises by rapidly fluctuating
discrete alternatives (on Planckian time scales). Here it is argued that
there is no conflict with quantum mechanics at larger distances
and time scales. Perhaps this is alternative path to 
resolving these issues, but reveals the radical revision required
to achieve consistency.  The Dvali-Gomez theory of quantum
black holes is more conservative \cite{DG}, preserving the familar
features of quantum mechanics and argues that
no UV completion of revision of quantum theory is necessary. In the 
Dvali-Gomez theory the no-hair theorem, Hawking radiation, etc, are emergent only
for large $N$ (quanta) black holes that approach a classical limit.

\section{Effective Field Theory}

 We introduce a real scalar field $B_0(x)$
to decribe Schwarzschild  mini-black holes as pointlike ``particles''
of the minimal mass $M_P$, 
which become the excitations of $B_0(x)$.  
In what follows we will neglect Hawking radiation.
We choose the Lagrangian of $B_0(x)$ to be
\beq
\label{cc}
L=\half\partial B_0 \partial B_0 -\frac{1}{2}M^2  B_0^2 +J B_0 +\Lambda
\eeq

Hawking proposed a virtual black hole vacuum (VBH)
\cite{Hawk}, in which   
 the vacuum is viewed 
 as consisting of Euclidean (instanton-like) loops
 of tiny black holes,  appearing and disappearing 
on time scales of order $M_P^{-1}$
(for a review, see \cite{Amelino}).\footnote{Hawking's rationale behind considering Euclidean spacetime 
may in part have been that the black hole 
loops are simply finite action instantonic field configurations, and Hawking radiation does not arise in Euclidean space.
Given the source $J$ there is no decay of the static VEV, $V$ in our Higgs phase. In any case, the
decay width $\Gamma$ can be significantly less than $M_P$ of a physical Planckian SBH of mass $M_P$ \cite{Page}.} 
This is connected in the literature
to various ideas in AdS hologaphy, gravitational instantons and
string theory \cite{Chamblin}\cite{Garay}\cite{Gibbons}\cite{Isham}.
The consequences of topological instanton fluctuations at
the Plank scale, 
associated with anomalous neutrino currents, have also been considered 
recently \cite{Dvali}.
Other authors have discussed possible gravitational effects in
neutrino physics \cite{Barenboim, Benatti}.

For a simple model of a VBH, 
we have added a source term, $J$.
The vacuum value of $B_0$ is therefore shifted
and we obtain the field $B_0 = B + V$ where $V= J/M^2$ in a Higgs phase:
\beq
L\rightarrow \frac{1}{2}\partial B \partial B -\frac{1}{2}M^2 B^2  -\frac{J^2}{2M^2} +\Lambda
\eeq
We can always choose the source term to cancel an anti-deSitter cosmological constant
$\Lambda= J^2/2M$.
When particles of the standard model propagate through a VBH
they interact with the $B+V$ field.
Our main objective presently
is to provide a field theory description of physical processes that involve matter interactions with
the SBH's.
This is immediately related to the issue of information loss as described in the introduction.

In the original
view of Hawking, global charges would simply be swallowed by mini-black holes
which subsequently evaporate, 
and large violations of global charge conservation would be expected. 
This has been implemented to conjecture, e.g.,  gravitationally induced  violation of
$B+L$ in the standard model \cite{Kane} (see also an exception, \cite{Stojkovic}). 
In the
modern prevailing view
global charges are conserved, with the global charge ``information'' holographically
painted onto the horizon of the black hole, to be recovered upon evaporation
\cite{holographic}.\footnote{We are not considering anomalies, 
as in \cite{Dvali}, which more definitively 
break the global symmetries by violating the conservation of the 
global current and  with an instanton
allows a mechanism to exchange visible charge with the vacuum.}

  For $N$ Weyl fermions we have the global $SU(N)\times U(1)$
kinetic term $\bar{\psi}^{\dot{\alpha}}_i i\partial_\mu \sigma^\mu_{\dot{\alpha}\beta} \psi^{i\beta}$.
For the Weyl fermion pair, $\epsilon^{\alpha \beta}\psi^i_\alpha \psi^j_\beta\sim [\psi^i \psi^j]$, we require 
 a complex spurion, $\theta_{ij}(x)$, to tie indices of fermions onto $B(x)$, as  
 \bea
 \label{inter}
 [\psi^i(x)\psi^j(x)]\theta_{ij}(x)\frac{B^2_0(x)}{M_P^2} + h.c.
 \eea
 For the case of $\psi$ representing neutrinos this is depicted in Fig.(3).
  Note, we include here $B^2_0(x)$ since the fermion pair is colliding with an existing
 black hole in the initial state and producing one in the  final state.\footnote{One could extend this to a model of production 
 with vertex $\sim g[\psi\psi]\theta B_0$ but this is complicated by a large energy dependence in 
 a form factor which suppresses $g(\mu)$ for $\mu << M_P$ \cite{BHP}.}
 In order to make an $SU(N)$ invariant interaction
 we require that $\theta$ lies in the symmetric representation of $SU(N)$
 of dimension $\half N(N+1)$. 
 
 Naturally, $\theta_{ij}(x)$ refers to the associated horizon information of $B_0(x)$.
 However, in the context of field theory we must face the issue of how to treat the spurion dynamically.
 In the following we consider two possibilities. 
 
 In case (II)  we argue that the spurion is simply a random complex valued variable
at the point of interaction.  We therefore average path integrals over $\theta_{ij}(x)$.
In this sense, the theory is analogous to a ``spin-glass'' in which spins propagate through
a random potential, and the partition function is averaged over the potentials \cite{Edwards}.
This promotes $\theta_{ij}(x)$ to a physical field.
Since gravity loses all memory of the information, there is no further autocorrelation
between $\theta_{ij}(x)$ and itself, e.g., no mass or kinetic terms, such as 
$\sim \theta_{ij}(x)\theta^{\dagger ij}(x)$.
Since there is no current associated with $\theta_{ij}(x)$. 
we interpret this as ``lost information.''

In case (III), following \cite{holographic}
we implement conservation of information, i.e., the holographic principle
which implies a conserved global current that involves the information.
To do this we must view the SBH as analogous to a very heavy atom, $H(x)$ (such as Uranium) that is bound to 
a light particle $\phi(x)$ (such as a neutron in its nucleus) 
to make a composite system, $\phi(x)H(x)$.  The physical properties
of the composite state are similar to those of the unbound $H(x)$, but the location in space-time
of the light field $\phi(x)$ must be correlated with the heavy $H(x)$.
However, the heavy particle carries the bulk mass of the system.  This imples both a conventional
kinetic term for $H(x)$, e.g., $\half \partial H\partial H$, and a 
conjoined kinetic term for the composite system,
\bea
\frac{1}{2M_P^2 }\partial_\mu(\phi(x)H(x))\partial^\mu(\phi(x)H(x)).
\eea
However, the mass term involves only $H(x)$, as $M^2 H(x)^2$.

In our case $H(x)\sim B_0(x)$ and $\phi(x) \sim \theta_{ij}(x)$. The conjoined kinetic term
$\partial(\theta^\dagger B_0)\partial(\theta B_0)$, and no
stand-alone $\partial(\theta^\dagger)\partial(\theta)$ term
implies
that $\theta(x)$ can never escape $B_0(x)$ unless the system decays through a fermion pair.
This leads to the following Lagrangian:
\bea
\label{KT}
L&=&\half\partial B_0 \partial B_0 -\frac{1}{2}M^2  B_0^2 +J B_0 +\Lambda
\nonumber \\
&& +\frac{1}{M_P^2}\partial(\theta^{*ij} B_0) \partial( \theta_{ij}B_0)
\eea

 The fact that $\theta$ is always
accompanied by a factor of $B_0(x)$ implies that it cannot escape the SBH, other than by
a neutrino interaction.
In the limit
$B_0\rightarrow V$ the kinetic term becomes:
\bea
\langle B_0\theta |L|B_0\theta \rangle & \sim & L_0 
+\frac{V^2}{M_P^2}Tr(\partial_\mu\theta^\dagger \partial^\mu \theta  )
\eea
Remarkably, in the VBH vacuum upon replacing $B_0\rightarrow V$ constant, 
the information field $\theta$  freely propagates through the medium. 
We can then absorb a factor of $V/M_P$ into $\theta$ to write
$(V^2/M^2)\partial_\mu(\theta^\dagger )\partial^\mu(\theta )\rightarrow
\partial_\mu\theta^\dagger \partial^\mu\theta$ and $\theta$ is
then canonical.
Mainly, there is now a conserved global current
that involves both the fermions and the field $\theta$.

The interaction vertex with the fermions then becomes, again,
that of eq.(\ref{inter}), in the VBH and suitably renormalized
\bea
 \label{inter2}
 [\psi^i(x)\psi^j(x)]\theta_{ij}(x) + h.c.
 \eea
The interaction annihilates an information-less SBH $B_0$ from the vacuum, and creates a composite SBH 
in the vacuum with information,
$\theta_{ij}(x)B_0(x)$.  $\theta$ is thus created (annihilated) by absorbing (producing) a fermion
pair, always in coincidence with $B_0$ through its VEV.  
We now apply this to sterile neutrinos.

 \section{Sterile Neutrinos}
 
\subsection{If Information is Lost}

Assume we have $N$ sterile neutrino flavors. This will imply an $SU(N)\times U(1)$
invariant kinetic term, and we assume this is a valid
global symmetry at the Planck scale.
Consider an $s$-wave pair of massless 
right-handed neutrinos scattering off of an SBH.
Here we have a unique situation in the SM that an $s$-wave combination of two massless right-handed fermions,
of flavors $i$ and $j$ can have zero local gauge coupling constant but nonzero global flavor. 

We assume the neutrinos
interact with the SBH $B_0$ field as in Fig.(1):
\beq
\label{above}
\nu_\alpha^i(x)\nu_\beta^j(x)\epsilon^{\alpha\beta}\theta_{ij}(x)
\frac{B^2_0(x)}{M^2_P}+h.c.
\eeq
where $\theta_{ij}(x)$ is   a constant  dimensionless spurion (note we
given $\theta$ dimensions of a mass). This is case(I) alluded to in the Intoduction and  will
break the conservation of the global currents of the neutrinos explicitly.
There is nothing intrinsic to gravity that can dictate
the flavor structure or phase of $\theta_{ij}$ in $SU(N)\times U(1)$
and we conclude that this is not a sensible theory.

\subsection{Random Field as a Spin-Glass}

A more reasonable possibility is (II) that $\theta_{ij}(x)$ is a complex random variable.
We have the interaction of eq.(\ref{above}), but $\theta_{ij}(x)$ can have no autocorrelation
due to gravity, since the SBH has lost all knowledge of the $ij$ indices,
hence no term like $\mu^2 \theta^*_{ij} \theta^{ij}$ is induced. 
This is the ``information lost'' scenario.  

A given choice of $\theta_{ij}(x)$ describes a particular subprocess.
However we then have to average over this field. The system is analogous to ``spin-glasses''
which have Hamiltonians that involve random variables 
(such as the Edwards-Anderson model \cite{Edwards}).

For spin-glasses
the averaging is done over the partition functions and not in the action itself.
In our case,  we average over the path integrals, and this
has the effect of promoting $\theta_{ij}$ to a random quantum field:
\bea
\label{path}
\rightarrow  \int D\theta \exp \left(i\int d^4x \;
\theta_{ij}\nu_\alpha^i\nu_\beta^j\epsilon^{\alpha\beta}(B_0^2/M_P^2)+h.c.\right)
\eea
While with  a fixed $\theta_{ij}(x)$
the neutrino flavor current conservation would be violated, 
it isn't hard to see that upon averaging over $\theta$
matrix elements yield a conserved global current 
$\langle \partial^\mu \bar{\nu}T^A\sigma_\mu \nu \rangle= 0$. 

The theory is singular, however, since the equation of motion of $\theta$ would enforce the
vanishing of the vertex. To see this, 
we pass to the VBH vacuum upon replacing $B_0\rightarrow V$ constant, 
and absorb the factor of $\sqrt{Z}=V/M_P$ into $\theta$ to canonically normalize $\theta$.
If we then introduce a small $\mu^2 \theta^*_{ij} \theta^{ij}$ term in the action, the
fermion current is manifestly conserved upon use of the equation of
motion of the neutrinos and $\theta$. 
In the VBH vacuum we have 
\beq
\label{4f}
\nu_\alpha^i\nu_\beta^j\epsilon^{\alpha\beta}\theta_{ij}
+h.c.-{\mu^2} \theta_{ij}\theta^{\dagger ij}
\eeq
where $Z=V^2/M_P^2$.
The corresponding 4-fermion interaction of eq.(\ref{4f}) is 
\beq
\frac{[\nu^i\nu^j][\bar\nu_i\bar \nu_j] }{ \mu^2}
\eeq
and is singular
as $\mu \rightarrow 0$. 
Likewise, as $\mu^2\rightarrow 0$ the equation of motion
of $\theta$  enforces $\nu_\alpha^i\nu_\beta^j\epsilon^{\alpha\beta}B_0(x)\rightarrow 0$.

The key feature  is the absence of the mass term for  $\theta_{ij}(x)$ in the Planck scale
effective Lagrangian.  
There are no derivatives of $\theta_{ij}(x)$ at this stage 
and no current built of $\theta_{ij}(x)$ and the theory is singular at $M_P$.
This would be in our opinion, a realization of
Hawkings information-lost hypothesis.  

We will see below, however, that this situation is unstable and  
effects of the back-reaction of the neutrinos will lead to a nonsingular dynamics
 for  $\theta_{ij}(x)$, on scales $\mu < M_P$ and a spontaneous breaking of $SU(N)$.
 The singularity at $M_P$ share features with a Landau pole.

\subsection{Information is Carried by Black Hole}

However, we can locally conserve the information (case III).  We  introduce a
new effective field, $B_{ij}= \theta_{ij}(x) B_0(x) $, that is composite 
and may represent a SBH
with the information of a neutrino pair encoded on its 
horizon holographically.
The effective theory for the neutrino pair interaction with
the BH's becomes:
\bea
\nu_\alpha^i(x)\nu_\beta^j(x)\epsilon^{\alpha\beta}
\frac{B_0(x)B^{ij\dagger}(x)}{M_P}+h.c.
\eea

How do we view $B^{ij}$?  An analogy was given in Section III
to the isotopes
of Uranium. We can freely add or remove neutrons from the Uranium 
nucleus, and the mass is not dramatically changed, nor are the chemical
properties.   Therefore we can view a Uranium atom as a groundstate nucleus, 
$U_0$, which with an additional neutron $n(x)$ becomes the effective field $U_0(x)n(x) $. 
Chemically (electromagnetically) we cannot 
easily discern which isotope we are dealing with.  
Yet another analogy might be bugs that end up flattened on the windshield
of a car, that have little effect on the properties of the car, 
but become part of a conjoined kinetic term with it.

Similarly, the effective field $B^{ij}(x)$ is essentially a
SBH, $B_0(x)$  with the information $\theta_{ij}(x)$ on it's horizon,
but there is no experiment we can do to detect $\theta_{ij}(x)$, other
than observing neutrinos emitted in Hawking radiation as the BH decays. The mass of a $\theta_{ij}(x)$
spurion field is zero.

The kinetic term becomes the ``conjoined kinetic term'' of eq.(\ref{KT}).
The mass term
is given by that of $B_0$ alone, $\half M_P^2 B_0^2$, 
and there can be no $Tr(\theta^\dagger\theta)$
term. 
The effective theory for the neutrino pair interaction 
in the VBH condensate becomes:
\bea
\nu_\alpha^i\nu_\beta^j\epsilon^{\alpha\beta}\theta_{ij}\frac{B_0^{2}}{M_P^2}+h.c.
\rightarrow 
Z\nu_\alpha^i\nu_\beta^j\epsilon^{\alpha\beta}\theta_{ij}+h.c.
\eea
where in the second term we have replaced $B_0$ by the condensate and $Z=V^2/M^2$.
Moreover, in the condensate the kinetic term becomes,
\bea
\partial B_{ij}^\dagger \partial B_{ij}
&\rightarrow  &
Z Tr(\partial_\mu\theta^\dagger \partial^\mu \theta  )
\eea
Therefore, if we canonically renormalize $\theta$ we can adjust $Z=1$
and our theory in the condensate becomes:
\bea
\label{start}
\nu_\alpha^i\nu_\beta^j\epsilon^{\alpha\beta}\theta_{ij}+h.c.+
\eta Tr(\partial_\mu\theta^\dagger \partial^\mu \theta  )
\eea
This provides an insight into what is meant by conservation or loss of information,
at least in the EFT:
If we conserve information then $\eta=1$; If information is (weakly) lost then $\eta =0$.
Note that with $\eta =1$ there is now a formal conserved information current
\bea
j^A_\mu =  \bar{\nu}T^A \sigma_\mu\nu+ i Tr(\theta^\dagger T^A\stackrel{\leftrightarrow}{\partial}_\mu \theta )
\eea
Information is now dynamically transferred from neutrinos to $\theta$ and  
can propagate freely through the condensate.

\begin{figure}[t!]
\vspace{-0.8 in}
	\hbox{\hspace{-0.8 in}\includegraphics[width=5 in]{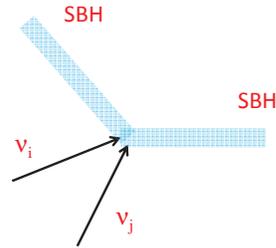}}
	\vspace{-1.0in}
	\caption{Low 4-momentum sterile neutrino pair of global flavors $(i,j)$ disappears into Schwarzschild BH (or the neutrino exchanges global charge with the SBH in a
	t-channel). This is described by a complex spurion field $\theta_{ij}$, and absence of an $M^2 Tr(\theta^*\theta) $. }
	\label{fig:vbhpd}
\end{figure}

\subsection{Back-Reaction of Neutrinos and Spontaneous Symmetry Breaking}

We can compute the action for the system
 at an energy scale $\mu$ by using the renormalization group.
This follows the procedure known as the ``block-spin renormalization group'' 
for treating the Nambu--Jona-Lasinio model \cite{NJL},
developed in Bardeen-Hill-Lindner \cite{BHL}.
For a fixed choice of $\theta_{ij}$
we integrate out the fermions, descending from a 
scale $ M_P$ to a scale $\mu$.  This is obtained from
the fermions loops  in Fig.(2) with loop-integrals,
\bea
\int_\mu^{M_P} \frac{d^4k}{(2\pi)^4}  .
\eea

\begin{figure}[t!]
\vspace{-2.2in}
	\hbox{\hspace{-1.9 in}\includegraphics[angle=90,origin=c,width=7 in]{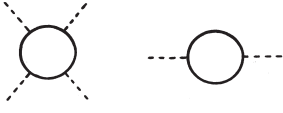}}
	\vspace{-3.0in}
	\caption{The infrared theory is controlled by the neutrino loops.}
	\label{fig:loops}
\end{figure}

We start with eq.(\ref{start}) as the defining action at $M_P$.
The result of integrating neutrino loops is an induced action for $\theta_{ij}$  of the form:
\bea
Z\partial_\mu \theta^*_{ij}\partial^\mu \theta_{ij}
-M^2_\mu\theta^*_{ij}\theta_{ij}-\frac{\lambda}{2} 
(\theta^*_{ij}\theta_{jk}\theta^*_{kl}\theta_{li})
\eea
Using the results for Weyl spinors in Hill, Luty and Paschos, \cite{HLP},
we immediately obtain:
\bea
Z &=& \eta+\frac{1}{8\pi^2}\ln\left(\frac{M_P^2}{\mu^2} \right)
\qquad \lambda = \frac{1}{\pi^2}\ln\left(\frac{M_P^2}{\mu^2} \right)
\nonumber \\
M_\mu^2 &=& -\frac{1}{4\pi^2}(M_P^2-\mu^2)
\eea 
Anticipating spontaneous symmetry breaking, we
denote  the field VEV, $\theta_{ij}=\delta_{ij}\theta$. Then
\beq
\theta^*_{ij}\theta_{ij}=N \theta^2 \qquad 
\theta^*_{ij}\theta_{jk}\theta^*_{kl}\theta_{li}=N\theta^4
\eeq
The unrenormalized potential is:
\bea
V_{un}=-\frac{N}{4\pi^2}(M_P^2-\mu^2)\theta^2+\frac{N\theta^4}{2\pi^2}\ln\left(\frac{M_P^2}{\mu^2} \right)
\eea
Note that the infrared cut-off on our loops is determined
by the neutrino mass. The unrenormalized neutrino mass
is $\mu = m_\nu = 2\theta$, and therefore the potential becomes
\bea
V_{un}=-\frac{N}{4\pi^2}M_P^2\theta^2+\frac{N\theta^4}{2\pi^2}
\left(\ln\left(\frac{M_P^2}{4\theta^2}\right) +2\right)
\eea
To renormalize we want the kinetic term to be brought
to canonical normalization.

The renormalized field VEV is 
therefore $\hat{\theta}^2=Z\theta^2$.
The neutrino mass becomes $\mu =m_\nu = 2 \hat{\theta}/\sqrt{Z}
=2g_r\hat{\theta}$, where $g_r=1/\sqrt{Z}$ is the renormalized 
Yukawa coupling. Note that if $\eta =0$ we see that $g_r$ displays the
 characteristic Landau pole at $\mu\rightarrow M_P$ which
is the compositeness condition of the field $\theta$ \cite{BHL}.
Thus we have the renormalized potential:
\bea
V_{ren}=-\frac{g_r^2N}{4\pi^2}(M_P^2)\hat\theta^2+\frac{g_r^4N\hat\theta^4}{2\pi^2}\left(\ln\left(\frac{M_P^2}{4g_r^2\hat\theta^2} \right)+2\right)
\eea
This is most conveniently rewritten in terms of the physical neutrino
mass:
\bea
V=-\frac{N}{16\pi^2}M_P^2m_\nu^2+\frac{Nm_\nu^4}{32\pi^2}\left(\ln\left(\frac{M_P^2}{m_\nu^2}\right) +2\right)
\eea
Here have substituted $m_\nu$ for the field $\theta$ and we
have a renormalization invariant potential as a function of a dynamical
$m_\nu$.

We now extremalize the potential with respect to $m_\nu$,
equivalent to extremalizing in $\hat{\theta}$.
We obtain:
\bea
0=
\frac{16\pi^2}{N}\frac{\partial}{\partial m_\nu^2}V=
-M_P^2 + {m_\nu^2}\left(\ln\left(\frac{M_P^2}{m_\nu^2}\right)+\frac{3}{2}\right)
\eea
The physical solution for the  neutrino mass is
\beq
m_\nu^2 = 0.424 \;M_P^2 \qquad m_\nu=0.651\; M_P
\eeq
The potential has a runaway for large values of $m_\nu>>M_P$, but this is
unphysical since we insist upon the cutoff $M_P$.

We remark that this result is sensitive to the subleading log
behavior of the loops (constants), which differs from \cite{BHL}. In that case
a large hierarchy is tuned by demanding a precisely tuned cancellation
between a bare mass term and the loop. Here we have no bare mass term
but we find a solution in a small log limit.
Hence, 
the boundstate field $\theta$
necessarily develop a vacuum expectation value (VEV) due to the VBH vacuum.

We note that we have neglected the production vertex, $\sim \nu\nu\theta B_0/M_P$,
since we are mainly interested in neutrino momenta below $M_P$. However, our
loop calculation informs us that the neutrinos, in the SBH Higgs phase, form a nonzero
VEV, $\langle \nu \nu \rangle \sim m_\nu M_P^2$ together with $\langle \theta \rangle \sim M_P$
which may be bootstrapped back to be
the source term $J$ for the BH condensate itself.  

\subsection{Phenomenology}

The sterile neutrinos thus obtain a large common Majorana mass,
$\sim M_P$.
The $N$ sterile neutrinos, coupled to gravity,
have a global $SU(N)\times U(1)$ symmetry which is now broken to $SO(N)$
 $\theta$ contains $N(N+1)$ real degrees of freedom.
The $SU(N)/SO(N)$ breaking implies there is one phase and  there remain
$\half N(N+1)$ massless Nambu-Goldstone bosons.

We consider the SM with 3 families including 3 sterile
neutrinos. The $SU(3)$ symmetry of the $\nu_{iR}$ is 
essentially an accidental symmetry given only the gravitational couplings of the neutrinos.
The left-handed lepton doublets, $\psi_{Li}$
couple via the Higgs boson to the right-handed neutrinos
through the Higgs field $H$
as:
\beq
y_{ij}\bar{\psi}_{Li} H\nu_{jR}
\eeq 
Generally the $y_{ij}$ will break the $SU(3)$.
Integrating out the heavy R-neutrinos yields the Weinberg operator,
\bea
\label{Wein}
\frac{1}{M_P}y_{ij}y^{kj}(\bar{\psi}_{Li} H )(\bar{\psi}_{Lk} H)^T +h.c.
\eea
If we now assume a typical (large) value for the $y^{kj}\sim 1$  in eq.(\ref{Wein}) we see that the 
scale of the induced 
Majorana mass terms of the observable L-neutrinos, with $v_{weak}\sim (175 \GeV)$,
is $\sim v_{weak}^2/(10^{19} \GeV) \sim 3\times 10^{-6}\eV$, which is rather small.
According to \cite{PDG}, the best fit to neutrino data implies we
require $\Delta m_{12}^2 \sim m_\nu^2 \sim 7\times 10^{-5} \eV^2$
which implies $m_\nu \sim 0.8\times 10^{-2} \eV$. Our results suggest a scale
of observable neutrino masses that is small by roughly a factor of $\sim 3 \times 10^{-3}$.

Our result for the Majorana mass scale depends only upon $M_P$ and is rather immutable.
However this is the Planck mass
at extremely high energies (of order  $M_P$).   It should be noted that
a number of authors have argued for significant renormalization effects of
the Planck scale, 
and that $M_P\sim 10^{16} \GeV$ may be reality in $D=4$ \cite{MP}.
Of course, with extra dimensions $M_P$ can be significantly
modified, but our set up requires  $D=4$ and would otherwise
have to be re-explored if $D\neq 4$.
However, neutrinos with the Type I seesaw may be uniquely probing 
gravity at $M_P$ and offer credence to a signficantly reduced
Planck mass at high energies.

The $\half N(N+1)=6$ Nambu-Goldstone bosons (Majorons) will have decay constant $f\sim M_P$
but potentials governed by the explicit $SU(3)$ symmetry breaking
Weinberg operator. Schematically $\sim m^4 \cos((\phi/f) +\chi )$, where $\chi$
is a CP phase and may range from $m\sim m_\nu$ to $m\sim v_{weak}$  depending upon the details of eq.(\ref{Wein}).
This potentially offers a number of cosmological possibilities, from late time phase
transitions, dark energy, to providing
an inflaton \cite{HS}. Discussion of these is beyond the scope of the present paper.

\section{Conclusions}

We have given an effective field theory treatment of 
information loss or conservation in the dynamics of a mini-black hole
interacting with fermions.  In particular, we have
focused upon
sterile neutrinos with a global $SU(N)\times U(1)$ symmetry.

Our present model illustrates how ``information'' might be described in analogy
to an induced effective random field $\theta$.  The weak information loss of global
charge would forbid  gravitationally induced auto-correlation of $\theta$, i.e., no kinetic, 
mass or interaction terms.  
However, consistency with the holographic view promotes $\theta_{ij}$
to a  local field and 
$\theta$ ``piggy-backs'' on a Schwarzshild black hole.
$\theta$ becomes dynamical in the black hole condensate.

The absence of the $\theta$ mass term for sterile neutrinos
 has an immediate and dramatic physical consequence.
 This implies that an instability driven by sterile fermion loops will always
lead to a condensate of $\theta\sim M_P$. In our case the instability is provided by the neutrino loops
external to the black holes, and $\theta$ becomes a sterile $\nu \nu$ boundstate.

Our present paper is introductory, but let us mention a future application.
In a subsequent paper  we  extend these results to locally
charged fermions.  We find that the dynamics is more
subtle. Owing to the local gauge field, the $B_{ij}$ field essentially describes a Reissner-Nordstrom (RN) black
hole. This acquires a slight mass enhancement of order $\alpha M_P$ above the mass
of the SBH $M_P$.  This mass enhancement is associated with the information field
$\theta_{ij}$, i.e., in the VBH the field $\theta$ freely propagates, but carries the charge of
the RN hole  and acquires the small RN mass term $\alpha M_P^2 Tr (\theta^\dagger \theta)$.
This means that the tachyonic instability is blocked 
for small $V/M_P$.  However, if an effective
coupling $g^2=V^2/M_P^2$ exceeds a critical value $g_c^2$ the field $\theta$ 
acquires a VEV and the gauge symmetry is then spontaneously broken.

A general picture that is emerging here, upon including local gauging, 
offers a new non-perturbative binding mechanism for fermions to produce scalar fields.
In turn, this suggests a large system of composite Higgs bosons.
Moreover, a near critical value of the coupling, $g\sim V/M$  implies deep scalar boundstates
 with (nearly) vanishing masses.
 Perhaps a more refined theory might lead to a conformal window with a low mass scale
 for the di-electron boundstate, $\theta$. In our crude approximations
 this would be a coupling tuned arbitrarily near criticality $g\approx g_c$. 
 Since there  are $1176$ Weyl bilinears
 in the standard model \cite{kids}, there may exist a large number of composite scalars in nature that are
 marginally subcritical boundstates of elementary fermions due to gravity, with masses
 that extend down to the Higgs mass scale $125$ GeV. In this picture, ``scalar democracy,''
 the standard model Higgs is composed of $\bar{t}t$, and its nearest neighbor
 $\bar{b}b$ would be expected at a mass scale $\sim 5.5$ TeV and accessible to
 an upgraded LHC \cite{kids}.  In this scheme, the Higgs boson(s) of the standard model are essentually mini-black holes \cite{CTHBH}

\section*{Acknowledgments}

We thank participants
of Simplicity-III at the Perimeter Institute, Waterloo, Canada, for discussions.
GB acknowledges support from the MEC and FEDER (EC) Grants SEV-2014-0398, 
FIS2015-72245-EXP, and  FPA2017-845438 and the Generalitat Valenciana under grant PROMETEOII/2017/033. 
She also acknowledges partial support from the European Union FP10ITN ELUSIVES (H2020-MSCAITN-2015-674896) 
and INVISIBLES-PLUS  (H2020-MSCA-RISE-2015-690575). CTH acknowledges
the  Fermi Research Alliance, LLC under Contract No.~DE-AC02-07CH11359 with the U.S.~Department of Energy, 
Office of Science, Office of High Energy Physics.

\end{document}